\documentclass[12pt]{article}
\usepackage{graphicx}
\usepackage{amsmath}
\usepackage{url}
\usepackage{courier}
\begin{document}
\begin{titlepage}
\title{{\bf Dalitz-plot dependence of $CP$ asymmetry in $B^{\pm} \to K^{\pm}K^+K^-$ decays}}
\author{
{L. Le\'sniak\footnote{email: leonard.lesniak@ifj.edu.pl} ~and P. \.Zenczykowski}\\
{\it Division of Theoretical Physics}\\
{\it The Henryk Niewodnicza\'nski
Institute of Nuclear Physics,}\\
{\it Polish Academy of Sciences}\\
{\it PL 31-342 Krak\'ow, Poland}
}
\maketitle
\begin{abstract}
A large $CP$-violating asymmetry predicted recently in the framework of the QCD factorization model for the 
$B^{\pm} \to K^{\pm}K^+K^-$ decays in the range of the $K^+K^-$ invariant masses above the $\phi(1020)$
resonance up to 1.4 GeV has been confirmed by the LHCb Collaboration.
We discuss the emergence and size of this asymmetry in an extended model involving 
$S$- and $P$-waves together with the contribution from the $D$-wave $f_2(1270)$ resonance
and compare our model with the experimental data of the LHCb and BABAR Collaborations. 
\end{abstract}
PACS numbers: 13.25.Hw, 13.75.Lb
\vfill

\end{titlepage}

Recently, the LHCb collaboration has measured the Dalitz-plot dependence of the $CP$-violating asymmetry in the
 three-body decay $B^{\pm} \to K^{\pm}K^+K^-$ \cite{LHCb}. The distribution of this asymmetry as
 a function of the $K^+K^-$ invariant mass squared (defined as
$m^2_{K^+K^-~{\rm low}} < m^2_{K^+K^-~{\rm high}}$ for two possible $K^+K^-$ combinations) has revealed that the observed $CP$ violation is not
 related to the $\phi(1020)$ resonance but is instead located in the region
 $1.2 < m^2_{K^+K^-~{\rm low}} < 2.0~{\rm GeV}^2$, with 
$m^2_{K^+K^-~{\rm high}} < 15~{\rm GeV}^2$ (see inset in Fig. 2(b) of Ref. \cite{LHCb}).
The measured $CP$ asymmetry turned out to be significantly 
   different from zero in the above range of $K^+K^-$ masses \cite{LHCb}:
$
A_{CP}(KKK)=-0.226 \pm 0.020 \pm 0.004 \pm 0.007,
$
where the first error is statistical, the second is systematic. and the third is due to the $CP$ asymmetry of the $B^{\pm}\to J/\psi K^{\pm}$ reference mode.
An earlier BABAR measurement of $CP$ asymmetry in a similar region suggested
its negative value as well \cite{BaBar}
(the three-body $B^{\pm}$ decays were also measured by Belle Collaboration \cite{Belle}).
Since  an even more negative estimate of the $A_{CP}(KKK)$ asymmetry was obtained  in the model presented in Refs. \cite{FKLZ,Lconf}, it is of some interest to discuss in this model the emergence and the $m^2_{K^+K^-~{\rm low}}$-- dependence of this asymmetry.

In Ref.~\cite{FKLZ} the $B^- \to K^-K^+K^-$ decay amplitude was calculated using the quasi-two-body QCD factorization
model. The matrix element of the effective weak Hamiltonian $H$ was expressed as:
\begin{equation}
\label{Amplit}
\langle K^-(p_1)K^+(p_2)K^-(p_3)|H|B^-\rangle = A^-_S+ A^-_P,
\end{equation}
where the $S$-wave part was given by
\begin{align}
A^-_S&=\frac{G_F}{\sqrt{2}}\left\{-\sqrt{\frac{1}{2}}\,\chi_S \,f_K (M^2_B-s_{23})
F_0^{B\to(K^+K^-)_S}(m^2_K)\,y\,\Gamma^{n*}_2(s_{23})\right. \nonumber \\
&\qquad \left. {}+
\frac{2B_0}{m_b-m_s}(M^2_B-m^2_K)F_0^{BK}(s_{23})\,\nu\,\Gamma^{s*}_2(s_{23})
\right\},
\label{AS}
\end{align}
and the $P$-wave part was
\begin{align}
A^-_P&=\frac{G_F}{\sqrt{2}}\left\{\frac{f_K}{f_{\rho}}\,A_0^{B\rho}(m^2_K)\,y\,
F_u^{K^+K^-}(s_{23})-
F_1^{BK}(s_{23})\left[w_uF_u^{K^+K^-}(s_{23})\right.\right. \nonumber \\
&\qquad \left.{}\left. +
w_dF_d^{K^+K^-}(s_{23})+w_sF_s^{K^+K^-}(s_{23})\right]
\right\}4\,\vec{p}_1\cdot \vec{p}_2.
\label{AP}
\end{align}

Above, the pair of interacting kaons of low invariant mass $s_{23}=m^2_{K^+K^-~{\rm low}}$ was defined as composed of kaons 2 and 3, while $\vec{p}_1$ and $\vec{p}_2$ are kaon 1 and kaon 2 momenta in the center-of-mass system of the kaons 2 and 3.
Furthermore, $G_F$ is the Fermi coupling constant,
$\chi_S$ is a constant related to the decay of the $(K^+K^-)_S$ state into two kaons, $f_K=0.1555$ GeV and $f_{\rho}=0.220$ GeV are the kaon and the $\rho$ meson decay constants,
$M_B$, $m_K$, $m_b=4.95$~GeV, $m_s=0.1$~GeV are the masses of the $B$ meson, kaon, $b$-quark, and strange quark, and $B_0 =m^2_{\pi}/(m_u+m_d)$, with $m_u$ and $m_d$ being the up and down quark masses. The constant $F_0^{B\to(K^+K^-)_S}(m^2_K)=0.13$ is the form factor of the transition from the $B$ meson to the $K^+K^-$ pair in the $S$-state \cite{ElBennich}, while functions
$\Gamma^n_2$ and $\Gamma^s_2$ are the kaon non-strange and strange scalar form factors.
Functions $F_0^{BK}(s_{23})$ and $F_1^{BK}(s_{23})$ are the $B\to K$ scalar and vector transition form factors \cite{FKLZ}. The constant $A_0^{B\rho}(m^2_K)=0.37 $ \cite{rhonumber} is the $B \to \rho$ transition form factor, while $F^{K^+K^-}_q$ ($q=u,d,s$) are kaon vector form factors \cite{FKLZ}. 
The short-range weak decay is governed by the QCD factorization coefficients and the Cabibbo-Kobayashi-Maskawa matrix elements that enter into expressions for $y$, $w_u$, $w_d$, $w_s$, and $\nu$.
For more detailed definitions, see Ref.~\cite{FKLZ}. The decay amplitude for the $B^+ \to K^+K^-K^+$ reaction is calculated in the way indicated in Ref. \cite{FKLZ}.

For the issue of the $CP$ asymmetry the form of the $P$-wave amplitude is less relevant than that of the $S$-wave one, since away from 
the $\phi(1020)$ resonance it is the $S$-wave that dominates the $K^+K^-$ effective mass spectra \cite{FKLZ}.
The main part of the asymmetry comes from the interplay of the strong phases of
the kaon non-strange and strange scalar form factors 
$\Gamma^n_2$ and $\Gamma^s_2$ with the weak amplitude phases hidden in $y$ and $\nu$.
The strong phases of $\Gamma^n_2$ and $\Gamma^s_2$ differ by more than $40^{\circ}$ at
$s_{23} = 1.0~{\rm GeV}^2$ with this phase difference diminishing only slowly for increasing $s_{23}$
(the difference vanishes at around $s_{23} = 2.0~{\rm GeV}^2$).
This behavior of the strong phases of $\Gamma^n_2$ and $\Gamma^s_2$ for the parametrization used in the present paper
is shown in Fig.~1. In Fig.~2 we give moduli of  the form factors $\Gamma^n_2$ and $\Gamma^s_2$.\\

\begin{figure}[h]
\begin{center}
\includegraphics[width=8cm,angle=270]{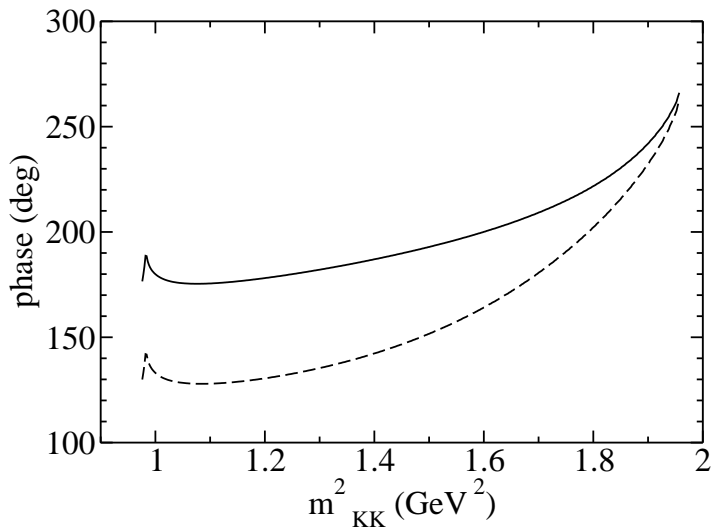}
\end{center}
\caption{Phases of the kaon non-strange and strange scalar form factors $\Gamma^n_2$ and $\Gamma^s_2$
(solid and dashed lines, respectively), calculated for the values of parameters $\kappa = 2.81 ~{\rm GeV}$, $c=0.109~{\rm GeV}^{-4}$, $f^s_2 = 0.7795~{\rm GeV}^{-2}$ introduced in Ref. \cite{FKLZ}.}
\end{figure}

\begin{figure}[h]
\begin{center}
\includegraphics[width=8cm,angle=270]{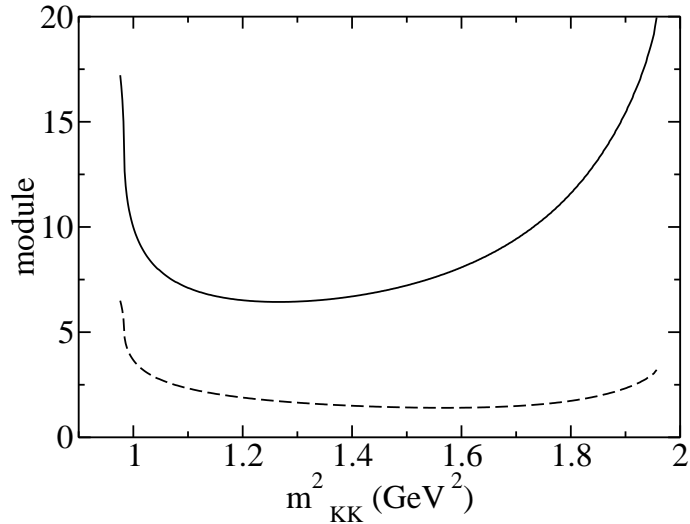}
\end{center}
\caption{Moduli of the kaon non-strange and strange scalar form factors $\Gamma^n_2$ and $\Gamma^s_2$ (solid and dashed lines, respectively),  calculated for the values of parameters $\kappa = 2.81 ~{\rm GeV}$, $c=0.109~{\rm GeV}^{-4}$, $f^s_2 = 0.7795~{\rm GeV}^{-2}$ introduced in Ref. \cite{FKLZ}.}
\end{figure}

 In addition to a large difference between the strong phases of $\Gamma^n_2$ and $\Gamma^s_2$
there is also a large difference
of about $-65^{\circ}$ between the phases of the weak amplitudes $y$ that are relevant for the $B^-$ and 
$B^+$ decays and are equal to
   -30 and +35 degrees, respectively.
The weak phase difference is mainly related to a presence of the Cabibbo-Kobayashi-Maskawa quark-mixing
 matrix element $V_{ub}$ depending on the 
angle $\gamma$ of the unitarity triangle (see Eqs. (6) and (13) of Ref. \cite{FKLZ}). On the other hand,
the phases of amplitudes $\nu$ are similar for the $B^-$ and $B^+$ decays and of the order of a few 
degrees only.  The presence of all these large strong and weak phase differences in
 the $K^+K^-$--mass region of interest has resulted in the large $CP$ violating asymmetry calculated in 
Ref.~\cite{FKLZ}, whose mass dependence was shown in Ref. \cite{Lconf}.
The plot presented in Ref.~\cite{Lconf} predicts the mass-dependent asymmetry considerably larger than that measured by the LHCb 
Collaboration \cite{LHCb}. It is therefore of some interest to inquire into the 
model dependence of our earlier estimates. 
One has to remember that the model of Ref. {\cite{FKLZ}} was fairly simplified.
In particular it did not take into account a possible effect of the $f_2(1270)$ resonance that could be important in the region of interest, i.e. for $1.2 < m^2_{K^+K^-~{\rm low}} < 2.0~{\rm GeV}^2$.\footnote{Although, as stated in Ref. \cite{FKLZ}, in the factorization approach the matrix element of weak current between the vacuum and the $K^+K^-$ state of total spin 2 does vanish, the transition amplitude between the latter state and the $B$ meson is in general non zero.} 
In the present paper, notwithstanding the small branching fraction of $f_2$ to $K\bar{K}$, the contribution of this resonance is included.
That is, the amplitude of Eq. (\ref{Amplit}) is supplemented with the $D$-wave contribution from the $f_2(1270)$ resonance. We calculate the relevant amplitude as 
for the $B^-\to \pi^-\pi^+\pi^-$ decay discussed
in Appendix A of Ref. \cite{DFKLL} (with pions replaced by kaons):
\begin{eqnarray}
A^-_D&=&\langle K^-(p_1)[K^+(p_2)K^-(p_3)]_D|H|B^-\rangle \nonumber \\
&=&-\langle f_2|u\bar{u}\rangle \frac{G_F}{\sqrt{2}} f_K\,G_{f_2K^+K^-}(s_{23})\,
F^{Bf_2}(m^2_K)\,y\,D(s_{12},s_{23}),
\label{AD}
\end{eqnarray}
where 
\begin{equation}
G_{f_2K^+K^-}(s_{23})=\frac{g_{f_2K^+K^-}}{m^2_{f_2}-s_{23}-im_{f_2}\Gamma_{f_2}}.
\end{equation}
The function $D(s_{12},s_{23})$ is given by Eq. (A.25) of Ref. \cite{DFKLL},
and the effective form factor $F^{Bf_2}(m^2_K)$ is treated as a free parameter.\footnote{The effective form factor for $B \to f_2$ transition is composed of three terms which are not well known (see Eq. (10a) in Ref. \cite{Koreans}). } 
The $f_2K^+K^-$ coupling constant is evaluated from
\begin{equation}
g_{f_2K^+K^-}=m_{f_2}\sqrt{\frac{60 \pi \Gamma_{f_2K^+K^-}}{q^5_{f_2}}},
\end{equation}
where $q_{f_2}$ is the kaon momentum in the $K^+K^-$ center-of-mass frame and $\Gamma_{f_2K^+K^-}=\frac{1}{2}\cdot 4.6\% \cdot \Gamma_{f_2}$, with $m_{f_2}$ and $\Gamma_{f_2}$ being the $f_2$ mass and its total width taken from Ref. \cite{PDG}.
Using the mixing angle $\alpha_T=(81\pm 1)^{\circ}$ relevant for the quark composition of $f_2(1270)$ (p. 201 in Ref. \cite{PDG}) one gets
\begin{equation}
\langle f_2|u\bar{u}\rangle = \frac{1}{\sqrt{2}}\sin \alpha_T = 0.698.
\end{equation}

We have fitted the effective $K^+K^-$ mass distributions from the data of Ref. \cite{LHCb},
corresponding to the region $1.0 < m^2_{K^+K^-~{\rm low}} < 1.9~{\rm GeV}^2$, with 
$m^2_{K^+K^-~{\rm high}} < 15~{\rm GeV}^2$. 
The parameters were as in Ref. \cite{FKLZ}:
$\chi_S$, $\kappa$ and $c$ (the latter two enter the scalar form factors $\Gamma^n_2$ and $\Gamma^s_2$),
 plus the new parameters $F^{Bf_2}(m^2_K)$
and the average $(m_u+m_d)/2$ of the light quark masses which are known not too well.
We minimize the function:
\begin{equation}
\label{chi2def}
\chi^2=\sum_{i=2}^{10}\left[
\left(\frac{N_i^{\rm th-}-N_i^{\rm exp-}}{\sigma^{\rm exp-}} \right)^2+
\left(\frac{N_i^{\rm th+}-N_i^{\rm exp+}}{\sigma^{\rm exp+}} \right)^2\,\,
\right],
\end{equation}
where $N_i^{\rm th\pm\,(\rm exp\pm)}$ are theoretical (experimental) numbers of events for $B^{\pm}$ decays
in nine bins of width $0.1~{\rm GeV}^2$ from $a=1.0$ to $b=1.9$~${\rm GeV}^2$ as measured by LHCb
\cite{LHCb} (the first data bin, i.e. 0.9--1.0~${\rm GeV}^2$ is omitted since it contains a region below 
the $K^+K^-$ threshold; we also assume that the corrections for acceptance do not affect the observed $CP$
asymmetry). 
The experimental errors are approximately equal to $\sigma^{\rm exp\pm}=\sqrt{N_i^{\rm exp\pm}}$.
We restrict our fit to $s_{23}<1.9~{\rm GeV}^2$ as  our knowledge of the KK interactions above 
         $s_{23}= 2$ GeV$^2$ is quite limited and furthermore our model includes an idealized $4\pi$ threshold at
 $(2 \cdot 0.7~{\rm GeV})^2=1.96~{\rm GeV}^2$.
As explained in Ref. \cite{KLL}, this threshold corresponds to the mass of the quasi-two-body system which effectively represents four pions coupled to the $K^+K^-$ pair. In reality, the threshold mass is diffused by interpion interactions.
The theoretical numbers of events are calculated from theoretical differential branching fractions 
of $B^{\pm}$ decays, i.e. from $\frac{dBr_i^{\pm}}{dm^2_{KK}}$, as
\begin{equation}
N_i^{\rm th\pm}=w_n \frac{dBr_i^{\pm}}{dm^2_{KK}},
\end{equation}
where
\begin{equation}
w_n=\frac{N^{\rm exp}}{\int_{a}^{b}dm^2_{KK}\left(\frac{dBr^{+}}{dm^2_{KK}}+\frac{dBr^{-}}{dm^2_{KK}}\right)}
\end{equation}
and
\begin{equation}
N^{\rm exp}=\sum_{i=2}^{10}(N_i^{\rm exp+}+N_i^{\rm exp-}).
\end{equation}

Since the $i=2$ bin involves a strongly varying contribution of $\phi(1020)$, the
theoretical values of $\frac{dBr_{i=2}^{\pm}}{dm^2_{KK}}$ are evaluated as integrals over the bin range. For the remaining bins, the theoretical branching fractions correspond to bin centers.

\begin{table}[h]
\caption{Model parameters fitted to the $K^+K^-$ effective mass distributions.}
\begin{center}
{\begin{tabular}{ccc} 
 &~~LHCb data - Ref. \cite{LHCb}~~ & ~~BABAR data - Ref. \cite{BaBar}~~ \rule{0mm}{6mm}\vspace{4pt}\\
\hline
 $ \chi_S~({\rm GeV}^{-1})$&
$ 6.83 \pm 0.99 $&$ 2.12 \pm 1.55 $ $\vphantom{\frac{1}{fg}}$\rule{0mm}{6mm}\\
 $ \kappa~({\rm GeV})$&
$ 2.81 \pm 0.29 $&$ 2.25 \pm 0.43 $ $\vphantom{\frac{1}{fg}}$\rule{0mm}{6mm}\\
$ c~({\rm GeV}^{-4})$&
$ 0.109 \pm 0.073 $&$ 0.069 \pm 0.173 $ $\vphantom{\frac{1}{fg}}$\rule{0mm}{6mm}\\
$ F^{Bf_2}(m^2_K)$&
$ 11.9 \pm 1.3 $&$ 12.3 \pm 3.5 $ $\vphantom{\frac{1}{fg}}$\rule{0mm}{6mm}\\
$ \frac{1}{2}(m_u+m_d)~({\rm MeV})$&
$ 2.74 \pm 0.26 $&$ 2.98 \pm 0.40 $ $\vphantom{\frac{1}{fg}}$\rule{0mm}{6mm}\vspace{4pt}\\
\hline
$\chi^2$&$34.5$&$18.0$$\vphantom{\frac{1}{fg}}$\rule{0mm}{6mm}\\
$ndf$&$13$&$11$ $\vphantom{\frac{1}{fg}}$\rule{0mm}{6mm}\\
$\chi^2/ndf$&$2.65$&$1.64$ $\vphantom{\frac{1}{fg}}$\rule{0mm}{6mm}\vspace{4pt}\\
\hline
\end{tabular}}
\end{center}
\label{fitparamet}
\end{table}

We have also fitted the effective $K^+K^-$ mass distributions shown in Fig. 8 of Ref. \cite{BaBar}, corresponding to the region $0.975 < m_{K^+K^-~{\rm low}} < 1.375~{\rm GeV}$.
We minimize the function $\chi^2$ analogous to Eq. (\ref{chi2def}) with the statistical errors $\sqrt{N_i^{\rm exp\mp}}$ replaced by the errors read from Fig. 8 in Ref. \cite{BaBar}. The fit has been performed for the eight bins of width $0.05$ GeV in the above region using the same parameters as for the LHCb case.

Our two fits to the 18 experimental LHCb data points and the 16 BABAR points yield the parameter values given in Table 1.
 The parameter errors have been rescaled by a factor
          $S=\sqrt{\chi^2/(ndf-1)}$ where $ndf$ is the number of degrees of freedom.

\begin{figure}[h]
\begin{center}
\vspace{10pt}
\includegraphics[width=10cm]{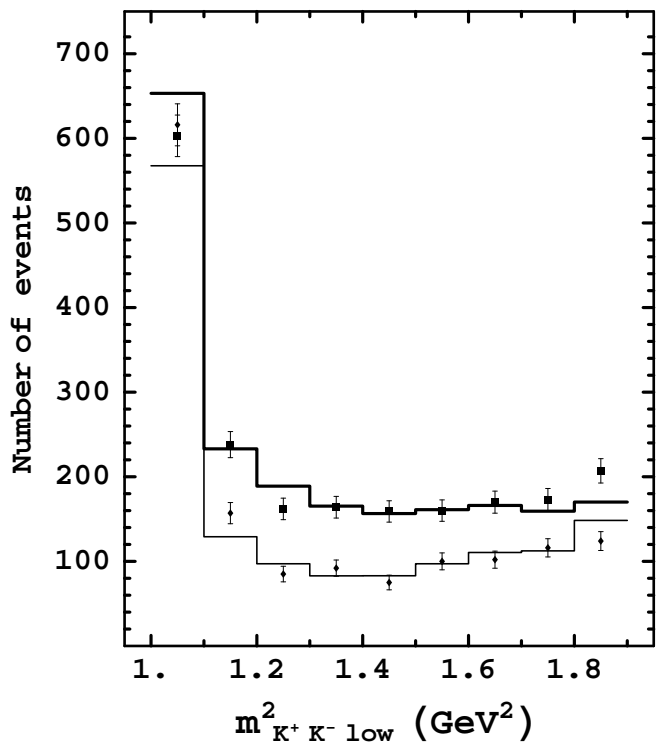}
\vspace{-10pt}
\end{center}
\caption{Numbers of the LHCb signal events for $B^{\pm}\to K^{\pm}K^-K^+$ decays as a function of $m^2_{K^+K^-~{\rm low}}$ in bins of $0.1~{\rm GeV}^2$: data  from Ref. \cite{LHCb}
($B^+$ -- squares, $B^-$ -- diamonds); our fit (Table \ref{fitparamet}) shown as thick ($B^+$) and thin ($B^-$) histograms.}
\label{events}
\end{figure}
\begin{figure}[t]
\begin{center}
\includegraphics[width=10cm, angle=0]{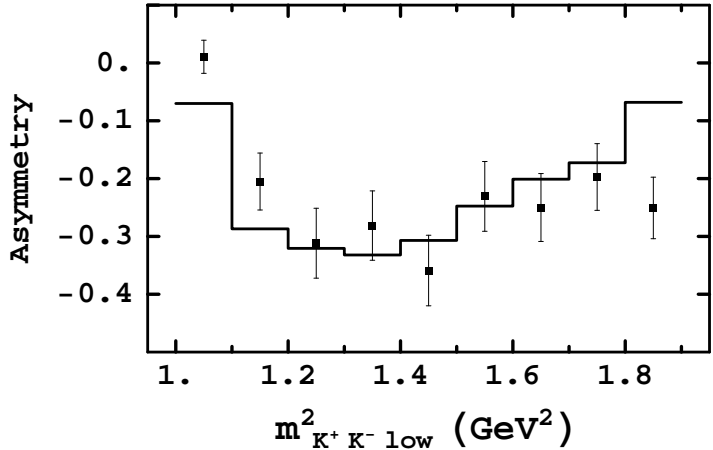}
\vspace{-15pt}
\end{center}
\caption{The $CP$-violating asymmetry $A_{CP}(KKK)$ as a function of $m^2_{K^+K^-~{\rm low}}$ in bins of $0.1~{\rm GeV}^2$: the LHCb data points from Ref. \cite{LHCb}; solid histogram --- the fit of Table \ref{fitparamet}.}
\label{asymmetry}
\end{figure}
In order to study the model dependence of our fits we decided to look at the model form factors $\Gamma^n_2$ and $\Gamma^s_2$. We have found that the largest uncertainty comes from the $f^s_2$ slope parameter 
 of the strange scalar form factor $\Gamma^s_2$ (Eqs. (16) and 
(20) of Ref. \cite{FKLZ}). Thus, 
for the LHCb data we have varied $f_2^s$
within its error limits coming from the numerical precision of the chiral 
perturbation theory constants $L^r_4$ and $L^r_5$. The value of $\chi^2=34.5$ corresponds to the upper value of
 $f^s_2=0.7795~{\rm GeV}^{-2}$, while for the central value of $f^s_2=0.6235~{\rm GeV}^{-2}$ used in 
Ref. \cite{FKLZ} we obtain a slightly worse fit with $\chi^2 = 39.9$.
Without the inclusion of the $f_2(1270)$ resonance the fit is significantly worse with $\chi^2 = 99.8$.
We do not include a contribution of the $f'_2(1525)$ resonance since its mass squared exceeds the upper limit of $1.9~{\rm GeV}^2$ chosen in our fit.
As in Ref. \cite{FKLZ}, the $P$-wave normalization constant $N_P$, common for the $B^-$ and $B^+$ decay 
amplitudes, was fixed at the value $N_P=1.037$. When the value of $F^{Bf_2}(m^2_K)$ is fixed at 11.93, then
 it is possible to make another fit with a free $N_P$ parameter and obtain essentially the same $\chi^2=34.5$ value with $N_P=1.0379\pm0.0615$.
Thus, one finds that $N_P \approx 1$ within its errors.
The obtained description of the LHCb data is shown in Figs. \ref{events} and \ref{asymmetry}
with the $CP$-violating asymmetry defined as
\begin{equation}
 A_{CP}(KKK)=\frac{\frac{dBr^{-}}{dm^2_{KK}}-\frac{dBr^{+}}{dm^2_{KK}}}
{\frac{dBr^{-}}{dm^2_{KK}}+\frac{dBr^{+}}{dm^2_{KK}}}.
\end{equation}
The first bin in Fig. 3 is dominated by the $\phi(1020)$ meson contribution and in this bin the $CP$-violating
asymmetry is small. 
In the next bins one observes that the number of events from the $B^{+}$ decays significantly exceeds 
the corresponding number of events for $B^-$ decays. This leads to a substantial negative asymmetry
shown in Fig. 4. In absence of the $D$-wave amplitude $A_D$ in the total decay amplitude the theoretical $CP$ asymmetry would be far more pronounced.

\begin{figure}[h]
\begin{center}
\includegraphics[width=10cm]{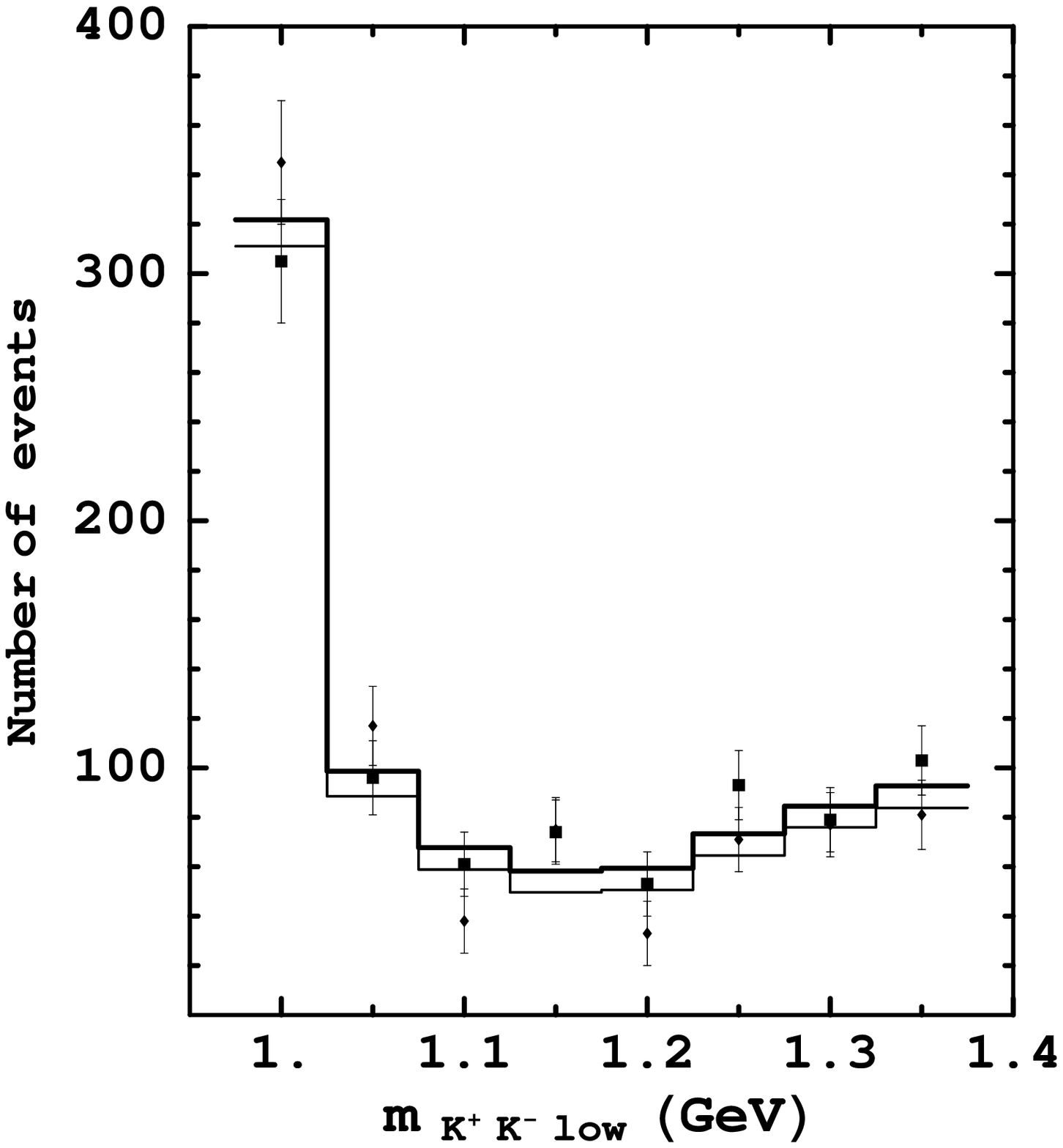}
\vspace{-10pt}
\end{center}
\caption{Numbers of the BABAR signal events for $B^{\pm}\to K^{\pm}K^-K^+$ decays as a function of $m_{K^+K^-~{\rm low}}$ in bins of $0.05~{\rm GeV}$: data  from Ref. \cite{BaBar}
($B^+$ -- squares, $B^-$ -- diamonds); our fit (Table \ref{fitparamet}) shown as thick ($B^+$) and thin ($B^-$) histograms.}
\label{BaBarevents}
\end{figure}
\begin{figure}[h]
\begin{center}
\includegraphics[width=10cm, angle=0]{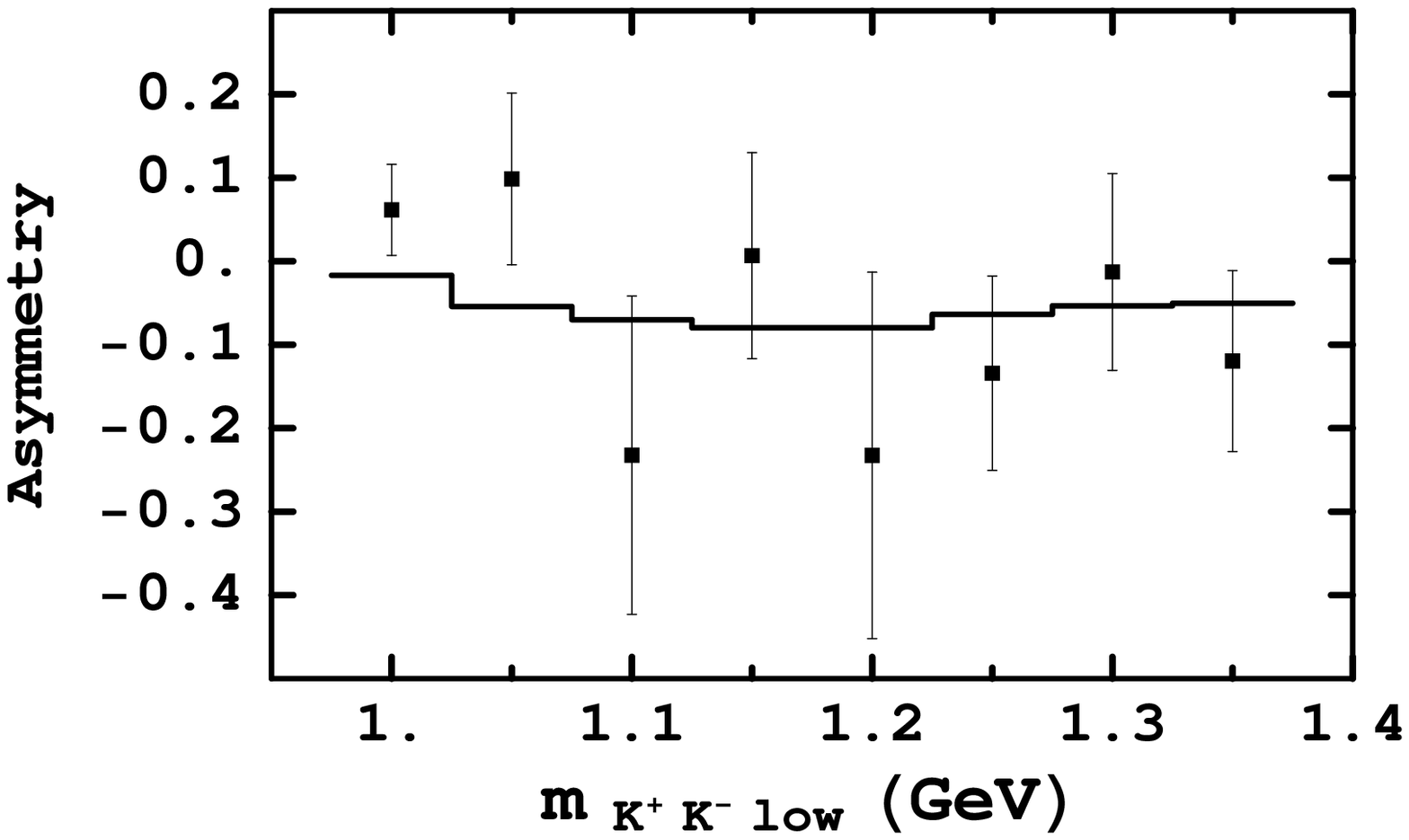}
\vspace{-15pt}
\end{center}
\caption{The $CP$-violating asymmetry $A_{CP}(KKK)$ as a function of $m_{K^+K^-~{\rm low}}$ in bins of $0.05~{\rm GeV}$: the BABAR data points from Ref. \cite{BaBar}; solid histogram --- the fit of Table \ref{fitparamet}.}
\label{BaBarasymmetry}
\end{figure}

For the BABAR data the corresponding distributions of the number of events and the $CP$ asymmetry are shown in Figs. \ref{BaBarevents} and \ref{BaBarasymmetry}. 
Note that the BABAR $CP$ asymmetry is significantly smaller than that found by the LHCb Collaboration. This difference is also stressed in Ref. \cite{BaBarLHCbComparison} where both data sets are compared.

Out of five parameters fitted to the LHCb and BABAR data the values of $\kappa$, $c$, $F^{Bf_2}(m^2_K)$ and $\frac{1}{2}(m_u+m_d)$ are equal within their errors while the value of $\chi_S$ is larger for the LHCb data (for the BABAR data we use the same value of $f^s_2=0.7795~{\rm GeV}^2$). The LHCb errors, taken as statistical only, are smaller than the BABAR data errors, whence larger value of $\chi^2$ for the LHCb case. The $CP$ asymmetry is roughly proportional to the constant $\chi_S$ (Eq. (\ref{AS})). As a result, the smaller experimental BABAR asymmetry  drives $\chi_S$ down (from the number of events outside of the $\phi(1020)$ region one gets the $CP$ asymmetry value $-(25.5 \pm 2.0)$\% for the LHCb data and $-(6.4 \pm 3.1)$\% for the BABAR data, where the errors are statistical).

In the $\phi(1020)$ region, we can well fit the BABAR data shown in Fig. 8 of Ref. \cite{BaBar} with the $\chi^2$ value $74.5$ for the 52 data points in the $m_{K^+K^-}$ region between $0.9925$ GeV and $1.060$ GeV. Since in this range the $P$-wave part of the decay amplitude dominates, one can fit only two $S$-wave parameters $\chi_S$ and $\frac{1}{2}(m_u+m_d)$. Their values appear consistent within large errors with the corresponding ones shown in Table \ref{fitparamet}.\\

We conclude that our model satisfactorily describes the size and the $K^+K^-$ mass dependence of the $CP$
 asymmetry in the region $(1.0<m^2_{K^+K^-~{\rm low}}<1.9)~{\rm GeV}^2$.
Both fits to the LHCb and BABAR data are improved when the  contribution of the $B^{\pm}$ decay amplitudes into the $K^{\pm}(f_2(1270)\to K^+K^-)$ final states is included.
\newpage

Acknowledgements\\

We would like to thank Irina Nasteva and Mariusz Witek for very useful
         discussions concerning the LHCb data presented in Fig. 2 of Ref. [1].

\vfill

\vfill
\end{document}